\documentclass[%
 reprint,
 amsmath,amssymb,
 aps,
]{revtex4-2}

\usepackage{graphicx}
\usepackage{dcolumn}
\usepackage{bm}

\begin{document}


\title{Diffusion in crowded colloids of particles cyclically changing their shapes}

\author{Yuki Koyano}
 
\affiliation{
Department of Physics, Tohoku University, 6-3, Aoba, Aramaki, Aoba-ku, Sendai 980-8578, Japan}

\author{Hiroyuki Kitahata}
\affiliation{
Department of Physics, Chiba University, 1-33 Yayoi-cho, Inage-ku, Chiba 263-8522, Japan}

\author{Alexander S. Mikhailov}
\affiliation{WPI Nano Life Science Institute, Kanazawa University, Kakuma-machi, Kanazawa 920-1192, Japan}
\affiliation{Abteilung Physikalische Chemie, Fritz-Haber-Institut der Max-Planck-Gesellschaft, Faradayweg 4-6, 14195 Berlin, Germany}

\begin{abstract}
A simple model of an active colloid consisting of dumbbell-shaped particles that cyclically change their length without propelling themselves is proposed and analyzed. At nanoscales, it represents an idealization for bacterial cytoplasm or for a biomembrane with active protein inclusions. Our numerical simulations demonstrate that non-equilibrium conformational activity of particles can strongly affect diffusion and structural relaxation: while a passive colloid behaves as a glass, it gets progressively fluidized when the activity is turned on. Qualitatively, this agrees with experimental results on optical tracking of probe particles in bacterial and yeast cells where metabolism-induced fluidization of cytoplasm was observed.
\end{abstract}

\maketitle

\date{\today}

\maketitle

\section{Introduction}
Typical examples of active colloids are provided by populations of microscopic particles or biological organisms that are able to propel themselves \cite{Gaspard}. It has been experimentally demonstrated that diffusion in the colonies of swimming bacteria can be strongly enhanced \cite{Libchaber_PRL_2000} and the theory for this effect is available too (see, e.g., \cite{Valeriani}). Active colloids can however be also formed at nanoscales by macromolecules that cyclically change their shapes under energy supply. Specifically, the cytoplasm of bacterial cells is known to represent a solution of conformationally active proteins, such as enzymes, motors and molecular machines \cite{Dey_Angew_2019}. This solution can be so crowded that the macromolecules almost touch one another within it. Moreover, biological membranes in a living cell typically include active protein inclusions that can make up about 40 \% of the membrane mass \cite{Franosch_Rep_2013}.

The estimates reveal that, in contrast to biological microorganisms, single active proteins cannot typically propel themselves \cite{footnote}: the conformational motions within their turnover cycles are only weakly non-reciprocal and the resulting propulsion forces are too small \cite{Sakaue,Alonso}. Therefore, such systems constitute a special class of active colloids, where individual particles are repeatedly changing their shapes, but do not swim. 

Even in absence of self-propulsion, persistent energy-driven conformational changes in macromolecules create non-thermal fluctuating hydrodynamical flows around them. Because passive tracer particles are advected by such fluctuating flows, their diffusion can become enhanced \cite{Mikhailov_PNAS_2015, KItahata_PRE_2016, Mikhailov_JPSJ_2018}. However, the analysis has so far been limited to dilute systems and therefore its results are not directly applicable to crowded colloids.

At sufficiently large volume ratios, passive colloids are known to have glass-like properties, manifested by slow relaxation, subdiffusion, and non-ergodicity (see, e.g., review \cite{Hunter_RepProgPhys_2012}). \textit{In vivo} experiments \cite{Parry_Cell_2014} on optical tracking of particles inside bacterial or yeast cells have been performed -- with a surprising result that, at least for the probe particles with relatively large sizes, such glass behavior is only characteristic under starvation conditions or in absence of the chemical fuel, such as ATP. The metabolism \textit{fluidizes} the cytoplasm, leading to the recovery of classical transport properties within it.

In this Letter, we propose an idealized model of oscillatory active colloids where the activity level and the rate of energy supply can be gradually controlled. The individual particles forming the considered colloid are active dimers, or dumbbells \cite{Kogler, Mikhailov_PNAS_2015}. It is known that such active dimers reproduce, in an approximate way, the mechanochemical activity in real enzymes and protein machines \cite{Mikhailov_Interface_2019}. Moreover, they have been already employed \cite{Dennison_SoftMatter_2017} in large-scale hydrodynamic simulations of the colloids (but still under less crowded conditions where glass effects are not yet seen). In contrast to the previous publications \cite{Mikhailov_PNAS_2015,Mikhailov_JPSJ_2018,Dennison_SoftMatter_2017,Mikhailov_Interface_2019}, our dumbbells are however further simplified: we assume that the natural length of the elastic spring that connects two beads in a dumbbell is periodically varied with time. The dumbbells interact via a soft parabolic repulsive potential and stochastic Langevin dynamics is assumed. In the present version, hydrodynamic interactions between the particles are dropped.

By running numerical simulations for relatively small 2D systems, we demonstrate that diffusion of dumbbells gets strongly enhanced when shape oscillations of its constituting particles are turned on. Further statistical investigations reveal that, while the passive colloid behaves as a glass, classical diffusion properties become recovered as conformational activity is increased.

\section{The model}

Enzymes are single-protein catalysts that convert substrate(s) into product(s) in each turnover cycle. In most enzymes, the cycles are accompanied by internal mechanochemical motions, i.e. by repeated changes in the conformation of a protein. Biological molecular motors and machines are also enzymes, with the only difference that mechanochemical motions are employed by them to manipulate other macromolecules or to produce mechanical work. To do this, chemical energy supplied with the substrate (often, ATP) is used. The bacterial cytoplasm or a cellular biomembrane are essentially densely packed colloids of active proteins that repeatedly change their shapes.

\begin{figure}[t]
  \centering
  \includegraphics{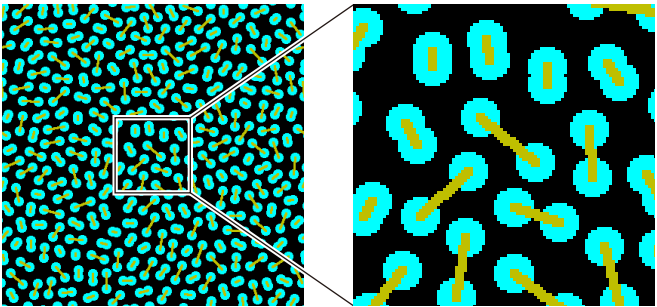}
  \caption{The crowded colloid of oscillating dumbbells. A snapshot from a simulation with $a_1=1$. The cyan circles, representing the beads, have radius $\sigma=0.68R$. Yellow lines symbolically show elastic strings connecting the beads in a dumbbell.
  }
  \label{fig1}
\end{figure}

Typically, mechanochemically active enzymes have a domain structure and their ligand-induced internal motions consist of the changes in distances between the domains and in the mutual orientation of them. Principal properties can be already reproduced in a simple active dimer model of an enzyme (see review \cite{Mikhailov_Interface_2019}). 

In the model, an enzyme protein is viewed as consisting of two beads (domains) connected by an elastic spring. The natural length of this spring depends on the ligand state of the enzyme: it is longer in absence of a ligand, but gets shorter after the substrate-enzyme and product-enzyme complexes are formed, returning to the original longer length after the product release. Thus, an enzyme behaves as a mechanochemical oscillator that undergoes an elongation and a contraction in each turnover cycle \cite{Mikhailov_Interface_2019}. 

This model has been successfully used to investigate collective hydrodynamic effects in numerical simulations for large populations of catalytically active enzymes \cite{Dennison_SoftMatter_2017}. In our study, it will be however further simplified: we will not explicitly consider the ligand states and chemical transitions between them. Instead, it shall be just assumed that the natural length of the spring connecting two beads in a dimer is periodically (with some additional random drift) changing with time. Physically, the modulation period should be considered as corresponding to the turnover time of an enzyme.

Explicitly, we assume that the natural length $\ell_i$ of dumbbell $i$ oscillates with time as 
\begin{equation}
\ell_i(t) = \ell_0 + \ell_1 \sin \psi_i(t),
\end{equation}
where the oscillation phase satisfies the equation
\begin{equation}\label{phase}
\frac{d \psi_i}{dt} = \omega_i + \zeta_i(t).
\end{equation}
Here, $\omega_i$ is the mean oscillation frequency and $\zeta_i(t)$ is the internal noise that takes into account stochastic variations in cycle times. This Gaussian noise is delta-correlated in time and independent for different dumbbells, 
\begin{equation}
\left< \zeta_i(t) \zeta_j(s) \right> = 2 \eta \delta_{ij} \delta(t-s).
\end{equation}
The parameter $\eta$ controls the characteristic coherence time for oscillations of the shape. In our numerical simulations, we assume that all dumbbells are identical and have the same oscillation frequency $\omega$. The model can however be readily extended to allow for random variation of these parameters.

Thus, the time-dependent elastic energy of dumbbell $i$ is
\begin{equation}\label{elastic}
    E_i(t)=\frac{1}{2}k \left (r_{12}^{(i)}-\ell_i(t) \right )^2,
\end{equation}
where $k$ is the stiffness of the internal spring and $r_{12}^{(i)}=\left| \bm{r}_i^{(1)} - \bm{r}_i^{(2)} \right|$ is the distance between the first and the second beads in the dumbbell $i$.

The considered colloid consists of $N$ dumbbells located within a volume of linear length $L$ (periodic boundary conditions are used). Beads belonging to different dumbbells interact via a soft repulsive potential 
\begin{equation}\label{parabolic}
u(r) = \left \{
\begin{array}{ll}
4u_0 \left( R - r \right)^2,&r \leq R \\
0, &r > R
\end{array}
\right .
\end{equation}
where $r$ is the half-distance between the beads. The parameter $u_0$ specifies the repulsion strength and $2R$ is the distance between the particles below which the repulsion starts. Note that there is no repulsion between the beads in the same dumbbell.

In molecular dynamics (MD) simulations of colloids, either full Newton dynamics or reduced stochastic dynamics can be used. However, at long times, both descriptions become equivalent \cite{Gleim_PRL_1998}. Below, stochastic Langevin dynamics will be employed. Furthermore, in the present version, possible hydrodynamic interactions between the particles are dropped, so that the effects of direct collisions are more clearly seen. They can however be introduced in the future into the model either by explicitly including the water molecules or in the framework of the multiparticle collision dynamics approximation (see, e.g., \cite{Dennison_SoftMatter_2017}).

Under the stochastic Langevin dynamics, the velocity of the bead $n=1,2$ in dumbbell $i$ is
\begin{equation}\label{Langevin}
\frac{d \bm{r}_i^{(n)}}{dt} = -\mu\frac{\partial E_i}{\partial \bm{r}_i^{(n)}}-\mu \frac{\partial U_i}{\partial\bm{r}_i^{(n)}}+ \bm{\xi}_i^{(n)}(t).
\end{equation}
Here $\mu$ is the mobility of the beads and
\begin{equation}
U_i \left( \bm{r}_i^{(n)} \right) = \sum_{j \neq i} \sum_{m=1,2} u \left(  \left|\bm{r}_j^{(m)} - \bm{r}_i^{(n)}\right|/2 \right)
\end{equation}
represents the potential experienced at position $\bm{r}_i^{(n)}$ by a bead $n$ of the dumbbell $i$, resulting from repulsive interactions with the beads of all other dumbbells.

Moreover, $\bm{\xi}_i^{(n)}(t)$ is the Gaussian thermal noise acting on the bead $n$ in the dumbbell $i$. Its correlation functions are
\begin{equation}
\left< \xi_{i,\alpha}^{(m)} (t) \xi_{j,\beta}^{(n)} (s) \right> = 2 \mu k_B T \delta_{mn} \delta_{ij} \delta_{\alpha \beta}\delta(t-s),
\end{equation}
where $T$ is the temperature, $k_B$ is the Boltzmann constant, and $\alpha,\beta=(x,y,z)$ in 3D or $\alpha,\beta=(x,y)$ in 2D.

It is convenient to measure all lengths in units of the repulsion interaction radius $R$ and the time in units of the relaxation time $\tau=(\mu k)^{-1}$ of the dumbbell. On the considered molecular scales, a convenient unit of energy is the thermal energy $k_BT$. 

After rescaling, the model is characterized by a set of dimensionless parameters: 
\begin{align}
    a_0 &=\frac{\ell_0}{R}, & a_1 &= \frac{\ell_1}{R}, & \Omega &=\omega \tau, \nonumber \\
    Y &=\eta \tau, & \kappa &=\frac{kR^2}{k_BT}, &\nu &=\frac{u_0 R^2}{k_BT}.
\end{align}

Through periodic modulation of the natural length of the spring connecting two beads, energy is persistently supplied to a dumbbell. To estimate its mean supply rate, we can notice that, in the steady state, it should be equal to the mean rate at which energy is dissipated by the dumbbell. A simple calculation yields that the energy $\Delta E$ supplied to an active dumbbell per one oscillation period is
\begin{equation}\label{supply}
    \frac{\Delta E}{k_BT}=\frac{ \pi a_1^2\kappa\Omega}{ 1+\Omega^2}.
\end{equation}
    
Additionally, an important parameter of the model is the fraction $\phi$ of the total volume occupied by the dumbbells.
Figure~\ref{fig1} shows an example of the simulated active colloids.

\section{The choice of parameters}
Depending on the parameter values, our model of an oscillatory colloid can describe various systems, including, for example, populations of biological microorganisms that cyclically change their shapes. The focus in the present study is however on the nanoscale phenomena within single living cells. Therefore, the parameter values typical for active proteins in biological cells shall be used.

Protein domains, corresponding to dumbbell beads in our model, typically have the size of about 10--20~nm. They are so stiff that one domain cannot deform another and penetrate inside it. Therefore, a repulsive hard-core potential could have been a good candidate to describe repulsive interactions between them. There are however also soft electrostatic interactions between proteins that extend over a few nanometers. Additionally, a protein is surrounded by a layer of hydrated water that is soft\cite{Chen_JPCB_2008}. Recently, non-contact effective interactions between proteins in water were analyzed by direct molecular dynamics simulations \cite{Feig_PCCP_2019}. Therefore, the hard repulsion core is effectively surrounded by a soft interaction shell.

\begin{figure}[t]
  \centering
  \includegraphics{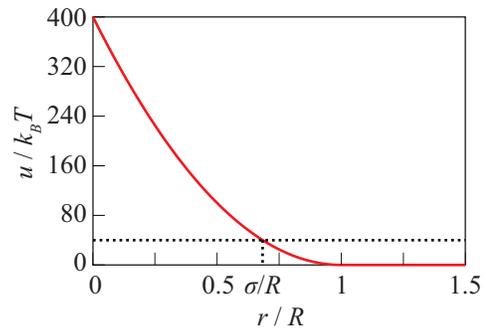}
  \caption{Parabolic repulsive interaction potential $u$ as a function of the half-distance $r$ between the beads from different dumbbells. For convenience, the length $\sigma=0.68R$ at which $u=40k_BT$ is chosen to define the ``hard-core'' radius of a bead.
  }
  \label{fig2}
\end{figure}

In our simple model, a single interaction potential (\ref{parabolic}) is used. Nonetheless, one can approximately interpret this potential as having a hard core and a soft outside shell. The boundary between them can be set at a radius $\sigma$ at which the repulsion potential becomes much larger than the thermal energy $k_BT$. Below, we take $u_0=100 k_BT/R^2$ in equation (\ref{parabolic}) and define the bead ``hard-core'' radius by the condition that $u(r=\sigma)=40 k_BT$, which yields approximately $\sigma=0.68 R$ (see Fig.~\ref{fig2}). Note however that such radius $\sigma$ is sensitive to the choice of the threshold potential value. It should be stressed that the model does not have a true hard-core repulsion potential and the beads can penetrate one into another to some extent.

If particles of linear size $\sigma$ are randomly distributed at the mean distance of $d$ between them, their volume fraction is about $\phi_{3D}= (\sigma/d)^3$ or the area fraction about $\phi_{2D}= (\sigma/d)^2$ in the 2D case, where the numerical prefactors that depend on particle shapes are dropped. 

In bacteria, about 30 percent of cytosol is typically occupied by proteins \cite{Vendeville}. Assuming that all proteins have the same size $\sigma$ and using the above estimate, this yields the relative distance of about $d/\sigma=0.3^{-1/3}\simeq 1.5$ between them. If, for example, $\sigma=0.68 R$, this corresponds to $d\simeq R\approx1.47\sigma$, so that the neighbours of a protein would typically be located within a soft interaction shell from it. A similar situation is characteristic for lipid membranes in biological cells. Typically, protein inclusions make up about 40 percent of the membrane mass and most of such inclusions (like, e.g., ion pumps) are active and cyclically change their shapes. Note that, in contrast to cytosol, the biomembrane represents a 2D colloid.

In this Letter, simulations for a two-dimensional system are performed. We shall have 246 dumbbells within the area with the linear size of $L=40 R$. Assuming that the radius of a bead is $\sigma=0.68 R$, this yields the area fraction of $\phi_{2D}=0.45$.

The relaxation time $\tau$ of the dumbbell should be about the characteristic slow conformational relaxation time in a protein, which is of the order of milliseconds. On the other hand, the turnover cycle time of an enzyme, corresponding in our model to the modulation period $2\pi/\omega$, typically takes tens of milliseconds. Therefore, we can choose $\Omega=0.1$. The initial oscillation phases will be randomly chosen for different dumbbells and noise with $Y = 10^{-1}$ will be moreover included into the phase evolution equation (\ref{phase}).  

When choosing the parameters $a_0$ and $a_1$, it is important that, even at the maximal length of a dumbbell, an additional bead cannot enter into the space between the beads within it, i.e. that the condition $(a_0+a_1)<4(\sigma/R)=2.72$ is satisfied. Below, we choose $a_0=1.5$ and vary $a_1$ between 0 and 1, so that this condition holds even at the largest oscillation amplitude.

The dimensionless stiffness of the spring connecting the beads is $\kappa=100$. The energy supplied to our model enzyme per single cycle is $\Delta E=7.8 k_BT$ at $a_1=0.5$ and $\Delta E=31.1 k_BT$ at $a_1=1$. For comparison, the chemical energy released in the reaction of ATP hydrolysis, often powering protein machines, is about $20 k_BT$.

\section{Numerical simulations}
In our two-dimensional simulations, periodic boundary conditions have been used. To prepare the initial configuration, the following procedure was employed: 

We started with a system where all dumbbells had a zero natural length, $\ell_0=0$, and they were regularly distributed forming a two-dimensional grid. Then, a short numerical simulation of this system over the time interval of $100\tau$ was performed. During this simulation, the system's temperature was raised 5-fold and, moreover, the natural length of dumbbell springs $\ell_0$, together with the modulation amplitude $\ell_1$, was gradually increased from zero to $\ell_0=1.5 R$ and $\ell_1 = a_1 R$. The equations were then integrated over further $100\tau$. Configurations of particles established at the end of such preparatory simulations were used as initial conditions for the actual simulations with the duration of $50000\tau$.

To explore the diffusion phenomena, we traced motion of the centers of mass, $\bm{\rho}_i = (\bm{r}^{(1)}_i + \bm{r}^{(2)}_i) / 2$, for all dumbbells $i$. Thus, trajectories were obtained that could be further analyzed. Averaging was always performed over all 246 different dumbbells in one simulation.

Note that at very short times, corresponding to displacements of mass centers much less than the mean distance between the particles, interactions between the dumbbells do not play a role. In this short-time regime, free Brownian motion of an isolated dumbbell, described by the Langevin equation (\ref{Langevin}) without the interaction terms, should be observed. For the considered system, the diffusion coefficient for an isolated dumbbell is $D_0=$0.005 $R^2/\tau$. At such very short times, the mean-square displacement (MSD) of a dumbbell from its initial position should be
\begin{equation}\label{free}
    \langle \Delta \bm{\rho}^2(t)\rangle=4 D_0 t.
\end{equation}

\begin{figure}[t]
  \centering
  \includegraphics{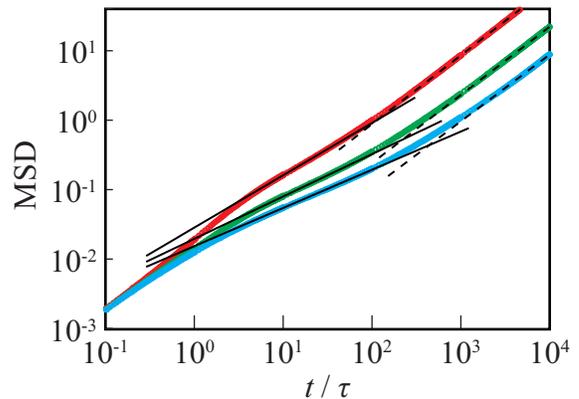}
  \caption{Dependences of the mean-square-displacement (MSD) of dumbbells (in units of $R^2$) on dimensionless time $t/\tau$ for systems with passive ($a_1=0$, blue) and active ($a_1=0.5$, green, and $a_1=1.0$, red) dumbbells. Solid and dashed lines show power-law fits with the exponents $\beta_-$ and $\beta_+$. The crossover times $t_{\text{cross}}$ correspond to intersections of these lines.  }
  \label{fig3}
\end{figure}

The mean-square-displacements of dumbbells, determined in our simulations, are shown as functions of time for different activity levels in Fig.~\ref{fig3}. 

It can be noticed that the dependences in this figure have the form characteristic for colloidal glasses \cite{Hunter_RepProgPhys_2012,Heuer}. As is usually done, they can be analyzed by fitting to power laws $t^\beta$. As is seen in Fig.~\ref{fig3}, one gets different exponents $\beta_-$ and $\beta_+$ in the intermediate- and long-time regimes. The crossover between the regimes occurs at times $t_{\text{cross}}$.
\begin{figure}
  \centering
  \includegraphics{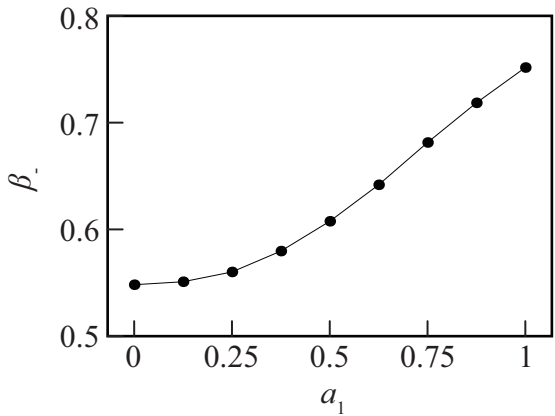}
  \caption{Dependence of the subdiffusion exponent $\beta_-$ on the activity level $a_1$. 
  }
  \label{fig4}
\end{figure}

\begin{figure}
  \centering
  \includegraphics{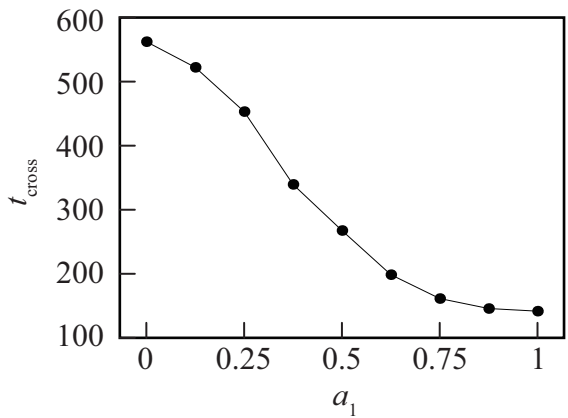}
  \caption{Dependence of the crossover time $t_\text{cross}$ on the activity level $a_1$. 
  }
  \label{fig5}  
 \end{figure} 

Both at short and at long times, classical diffusion is observed (exponents $\beta_+$ vary between 0.97 at $a_1=0$ and 0.99 at $a_1=1$). In the intermediate regime, subdiffusion with the exponents $\beta_-$ ranging between 0.55 and 0.75 takes place. The dependences of the subdiffusion exponent $\beta_-$ and the crossover time  $t_{\text{cross}}$ on the activity level $a_1$ are shown in Figs.~\ref{fig4} and \ref{fig5}.

While classical diffusion is again recovered at long times, the diffusion coefficient $D$ at such times is however smaller than $D_0$, indicating that diffusion in the colloid becomes suppressed.

Both the suppression of diffusion at long times and the observed subdiffusion at intermediate times are typical glass effects. They suggest that \textit{caging} of particles takes place \cite{Hunter_RepProgPhys_2012}. The free diffusive motion of a particle is blocked by other particles surrounding it and forming a cage. Displacements over large distances are only possible if, by a rare fluctuation, the particle was able to escape from its cage. 

\begin{figure}[t]
  \centering
  \includegraphics{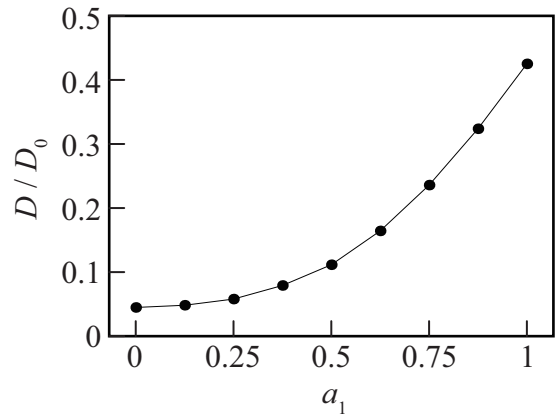}
  \caption{Dependence of the long-time diffusion coefficient $D$ of dumbbells on their activity level $a_1$.}
  \label{fig6}
\end{figure}

The dependence of the long-time diffusion coefficient of particles on the activity level $a_1$ is shown in Fig.~\ref{fig6}. We see that diffusion becomes enhanced when non-equilibrium conformational activity of dumbbells is introduced and gradually increased. Alternatively, it can be said that suppression of diffusion with respect to that for free particles becomes then less strong. Furthermore, as is seen in Figs.~\ref{fig4} and \ref{fig5}, the subdiffusion exponent $\beta_-$ gets larger at higher activity levels and the classical diffusion regime sets on earlier (i.e., at the shorter cross-over times $t_{\text{cross}}$) with an increase in the parameter $a_1$; it becomes then close to the oscillation period $T=2\pi/\Omega=62.8$ of active dumbbells.

Similar behavior is observed in equilibrium colloids when the volume or area fractions of particles are decreased; it corresponds to a transition from the glass to the fluid phase \cite{Hunter_RepProgPhys_2012,Heuer}. The above results suggest that effective fluidization of a colloid can also take place because of the non-equilibrium conformational activity of the particles forming it.

To further explore this conjecture, structural relaxation in the model has been numerically analyzed. To do this, we have determined the scattering function \begin{equation}
    F_2(k,t)=\frac{1}{N}\Big\langle\sum_i e^{i\bm{k} \cdot (\bm{\rho}_i(t)-\bm{\rho}_i(0))}\Big\rangle.
\end{equation}
Because of the isotropy of the system, this function depends only on $k=|\bm{k}|.$
\begin{figure}[t]
  \centering
  \includegraphics{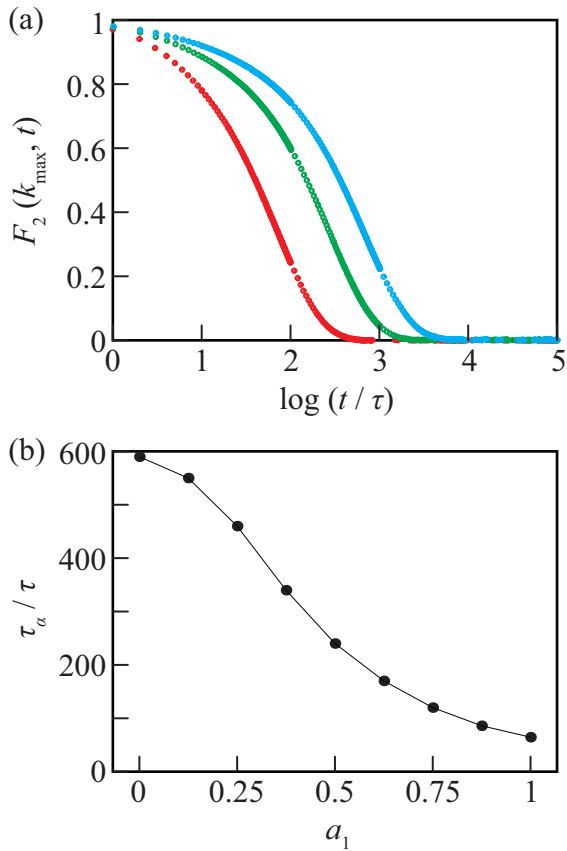}
  \caption{(a) Scattering functions $F_2(k_{\mathrm{max}},t)$ at $a_1=0$ (blue), 0.5 (green), and 1.0 (red). (b) Structural relaxation times $\tau_\alpha$ at various activity levels $a_1$.}
  \label{fig7}
\end{figure}

The determined functions $F_2(k_{\mathrm{max}},t)$ are shown for three different activity levels in Fig.~\ref{fig7}a. Here, $k_{\mathrm{max}}=2\pi/b$ where $b=L/N^{1/2}$ is the mean distance between the dumbbells. The structural relaxation time $\tau_\alpha$ is defined \cite{Hunter_RepProgPhys_2012} by the equation
\begin{equation}
    F_2(k_{\mathrm{max}},\tau_\alpha)=\frac{1}{e}.
\end{equation}
The dependence of $\tau_\alpha$ on the activity level $a_1$ is displayed in Fig.~\ref{fig7}b.

As we can see, structural relaxation gets much faster when non-equilibrium conformational activity in the particles takes place. The structural relaxation time decreases by an order of magnitude in comparison to the passive colloid ($a_1=0$) when $a_1=1$.

To further explore statistical properties of trajectories, we determined statistical distributions of displacements $\left|\Delta \bm{\rho}\right|$ within a given time $\Delta t$. For classical diffusion, such distributions should be
\begin{equation}
   p(\left| \Delta \bm{\rho} \right|)\varpropto \left| \Delta \bm{\rho} \right|\exp \left (-\frac{\left| \Delta \bm{\rho} \right|^2}{2 D t} \right ). 
\end{equation}

Therefore, the ratio  $p(\left| \Delta \bm{\rho} \right|)/\left| \Delta \bm{\rho} \right|$ has then the Gaussian form. 

\begin{figure}[t]
  \centering
  \includegraphics{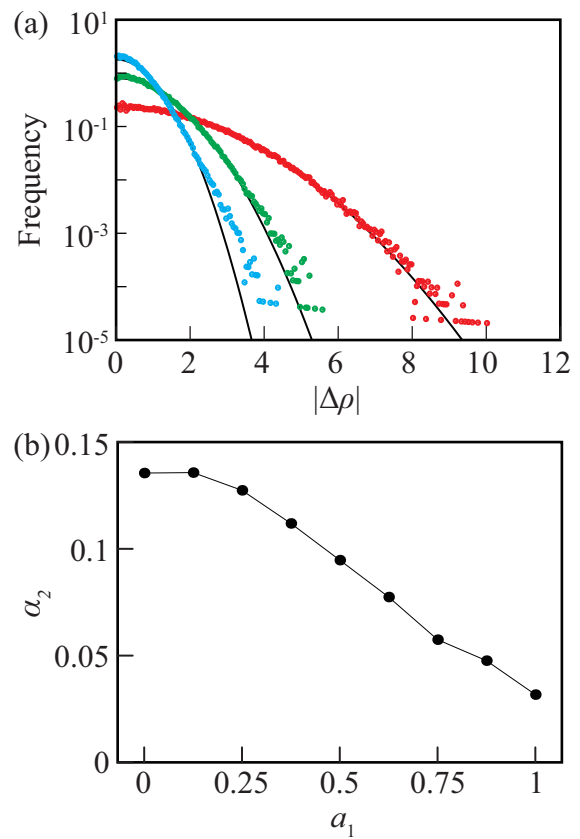}
  \caption{(a) Normalized distributions of displacements (see the text) within the time interval $\Delta t=1000\tau$ for passive ($a_1=0$, blue) and active ($a_1=0.5$, green, and $a_1=1.0$, red) dumbbells. Solid curves are fits to the Gaussian form. (b) Non-Gaussianity coefficients $\alpha_2$ at different activity levels $a_1$ for $\Delta t=1000\tau$.}
  \label{fig8}
\end{figure}

In Fig.~\ref{fig8}(a), we show the normalized histograms where the frequencies of displacements at three activity levels within time $\Delta t=1000 \tau$ are divided by $\left| \Delta \bm{\rho} \right|$. Additionally, this figure shows the fits of such distributions to the Gaussian form. It can be noticed that, for a passive colloid, the distribution deviates from the Gaussian dependence for large displacements $\Delta \bm{\rho}$. When non-equilibrium conformational activity is switched on, the deviations become smaller and they practically disappear at $a_1=1$.

To quantitatively characterize the deviations, the non-Gaussianity coefficient \begin{equation}
    \alpha_d= \frac{\left< \left|\Delta \bm{\rho} \right|^4 \right> }{ \zeta_d\left<\left| \Delta \bm{\rho}\right|^2 \right>^2}-1
\end{equation}
is introduced, where $\zeta_d=1+2/d$ and $d$ is the dimensionality of the system. This coefficient vanishes in the Gaussian case. 

The computed non-Gaussianity coefficients at different activity levels are shown in Fig.~\ref{fig8}(b). One can see that such coefficients decrease at higher activity levels, thus further supporting our conclusion that fluidization of the system takes place.

\begin{figure}[hb]
  \centering
  \includegraphics{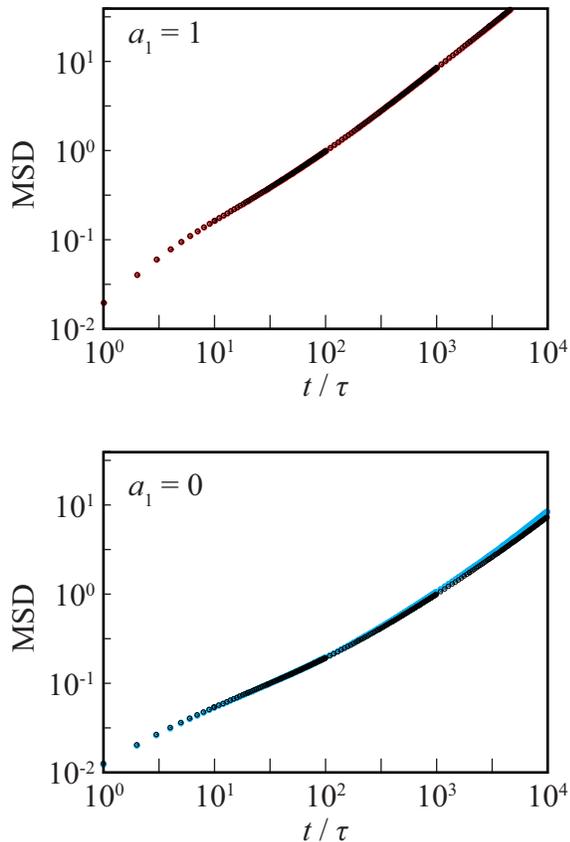}
  \caption{Dependences of MSD (in units of $R^2$) on dimensionless time $t/\tau$ for the system of size $L=40R$ with $N=246$ and the system of double size $L=80R$ with $N=988$ at $a_1=0$ (below)  and $a_1=1$ (above). Red and blue symbols are for the system of size $L=40R$, while black symbols are for the system of size $L=80R$.
  }
  \label{fig9}  
\end{figure} 

To check for possible finite-size effects, some simulations have been repeated for a system of the double linear size. Figure~\ref{fig9} demonstrates that the computed time dependences of MSD for the single- and double-size systems practically coincide.

In another test, we have taken as an initial condition a snapshot from the oscillatory system with $a_1=1$. Then, the simulation was continued, but with the natural lengths $\ell_i$ of dumbbells \textit{frozen} at their values in such initial snapshot. Thus, an equilibrium system was constructed with exactly the same, but static, size distribution of dumbbells as in the oscillatory case.

\begin{figure}
  \centering
  \includegraphics{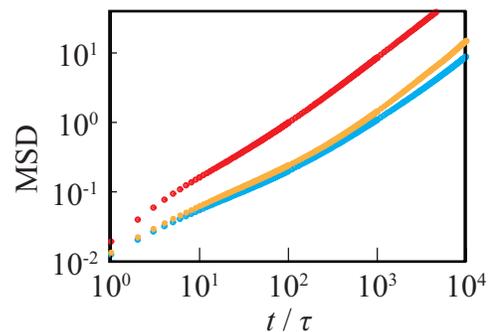}
  \caption{Dependences of MSD (in units of $R^2$) on dimensionless time $t/\tau$ for the passive colloid with $a_1=0$ (blue), the oscillatory colloid with $a_1=1$ (red) and the frozen colloid (orange). The frozen colloid is constructed by fixing the natural lengths $\ell_i$ of all dumbbells at their instantaneous values at some time moment in the oscillatory case with $a_1=1$. It represents a passive colloid with the same shape distribution of dumbbells as in the oscillatory case.
  }
  \label{fig10}
\end{figure}

The dependence of MSD on time in the frozen system is shown in Fig.~\ref{fig10}. From this dependence, we could find that the long-time diffusion constant is $D_{\text{frozen}}/D_0=0.072$. This value is closer to the diffusion constant $D/D_0=0.045$ in the passive colloid and much less than the diffusion constant $D/D_0=0.45$ in the respective oscillatory case. Hence, we have demonstrated that active oscillations, not the dispersion of sizes or possible overlaps, are responsible for the observed diffusion enhancement and for the fluidization of a colloidal glass.

\section{Discussion}

In this Letter, we have proposed a simple model of a non-equilibrium colloid of active particles reciprocally changing their shapes. The parameters of the model could be chosen in such a way that the conditions typical for metabolically active bacterial cytoplasm and for cellular membranes with active protein inclusions were roughly reproduced.

Colloids of circular particles, whose radii periodically changed, were previously considered \cite{Berthier} as a model for epithelial cellular tissue. This model is not applicable to protein colloids since, due to incompressibility of water and lipid bilayers, the volume of a particle should be conserved. Instead of the glass behavior, an analog of the yielding transition in amorphous solids was observed \cite{Berthier}.

In contrast, our simulations have demonstrated that passive colloids of dumbbells behave as glasses, but the glass properties fade away when conformational oscillations are introduced. They were performed for 2D systems, but, as typical for crowded colloids \cite{Hunter_RepProgPhys_2012,Heuer}, similar behavior can be expected in the 3D case too. Thus, we conclude that, under sufficiently strong conformational activity of the particles, effective fluidization of a glass-like colloid may take place. The fluidization becomes possible at the rate of energy supply of about 10 $k_BT $ per a particle and a turnover cycle.

The experimental effects depended however strongly on the size of the probe particles, getting less pronounced for the tracers of a smaller size \cite{Parry_Cell_2014}.  In our study, additional probe particles were not introduced and only trajectories of small dumbbells themselves were tracked. Therefore, our results cannot be directly compared with them. Nonetheless, some comments can be made.

According to Ref.~\cite{Ooshida}, cages in colloidal glasses are characterized by a hierarchical onion-like structure, with the smaller ones enclosed within the larger ones. If a smaller cage is destroyed and a particle escapes from it, it may still stay confined within a larger cage. Moreover, characteristic times of diffusion processes get progressively increased with the cage size, following a self-similarity law \cite{Ooshida}. It seems natural to assume that a large probe particle can be only confined within the cages of the respective large size. Moreover, one might also expect that diffusion for probe particles would be qualitatively the same as for the smaller particles, though scaled up in time. The cross-over from subdiffusion to classical diffusion would occur at MCD about the probe particle size. In the experiments \cite{Parry_Cell_2014}, only displacements at relatively long times could be resolved. For the smallest probe particles, the observation times could have been shorter than $t_{\text{cross}}$, so that the glass-like effects indeed remained weak.

For glass-like colloids, it is known that their statistical single-particle properties can be already reproduced using small systems, whereas many-particle correlations should have strong size effects\cite{Heuer} (see also \cite{Hunter_RepProgPhys_2012, Ooshida}). Thus, we could reliably investigate diffusion of individual particles in the present study with a small system, but further investigations aimed, e.g., at exploring dynamic heterogeneity effects, shall require working with the systems of a larger size. In the future, simulations accounting for hydrodynamic interactions can also be performed.

Based on our results, strong diffusion enhancement can be expected not only in bacterial cytoplasm, but also for biological membranes with protein inclusions, provided that chemical energy needed to maintain shape oscillations in membrane proteins is persistently supplied. It would be interesting to experimentally check this.

The observed fluidization of a colloidal glass can be interpreted as an effect of an additional non-thermal noise caused by active conformational changes in the particles constituting it. Hence, our study supports the conjecture \cite{Guo} (see also \cite{Yasuda}) that, in presence of metabolism, active intracellular noise might prevail over thermal noise and determine transport phenomena in the cells.

At the end, we want to stress that, although the discussion in this Letter was centered on the nanoscale processes within biological cells, the proposed model is universal; it can be applied to other, natural or artificial, systems at different length and time scales as well. Indeed,  artificial non-equilibrium colloids of small oscillating dumbbell-shaped particles can be readily designed.

\acknowledgments
Stimulating discussions with T. Ooshida and S. Komura are gratefully acknowledged. This work was supported by JSPS KAKENHI Grants No. JP19K03765 and No. JP19J00365 in Japan.


\begin{thebibliography}{99}

\bibitem{Gaspard}
P.~Gaspard and R.~Kapral, \textit{Adv. Phys. X} \textbf{4}, 1602480 (2009).

\bibitem{Libchaber_PRL_2000}
X.-L.~Wu and A.~Libchaber, \textit{Phys. Rev. Lett.} \textbf{84}, 3017 (2000).

\bibitem{Valeriani}
C.~Valeriani, M.~Li, J.~Novosel, J.~Arlt, and D.~Marenduzzo, \textit{Soft Matter} \textbf{7}, 5228 (2011).

\bibitem{Dey_Angew_2019}
K.~Dey, \textit{Angew. Chem., Int. Ed.} \textbf{58}, 2208 (2019).

\bibitem{Franosch_Rep_2013}
F.~H\"{o}fling and T.~Franosch, \textit{Rep. Prog. Phys.} \textbf{76}, 046602 (2013).


\bibitem{footnote} Self-propulsion might however take place for some exceptionally rapid enzymes \cite{Granick_PNAS_2018}.

\bibitem{Granick_PNAS_2018}
A.-Y.~Jee, S.~Dutta, Y.-K.~Cho, T.~Tlusty, and S.~Granick, \textit{Proc. Natl. Acad. Sci. USA} \textbf{115}, 14 (2018).

\bibitem{Sakaue}
T.~Sakaue, R.~Kapral, and A.~S.~Mikhailov, \textrm{Eur. Phys. J. B} \textbf{75} 381 (2010).

\bibitem{Alonso}
S.~Alonso and A.~S.~Mikhailov, \textit{Phys. Rev. E} \textbf{79} 061906 (2009).

\bibitem{Mikhailov_PNAS_2015}
A.~S.~Mikhailov and R.~Kapral, \textit{Proc. Natl. Acad. Sci. USA} \textbf{112} E3639 (2015).

\bibitem{KItahata_PRE_2016}
Y.~Koyano, H.~Kitahata, and A.~S.~Mikhailov, \textit{Phys. Rev. E} \textbf{94}, 022416 (2016).

\bibitem{Mikhailov_JPSJ_2018}
A.~S.~Mikhailov, Y.~Koyano, and H.~Kitahata, \textit{J. Phys. Soc. Jpn.} \textbf{86}, 101013 (2017).

\bibitem{Hunter_RepProgPhys_2012}
G.~L.~Hunter and E.~R.~Weeks, \textit{Rep. Prog. Phys.} \textbf{75}, 066501 (2012).

\bibitem{Parry_Cell_2014}
B.~R.~Parry, I.~V.~Surovtsev, M.~T.~Cabeen, C.~S.~O'Hern, C.~S.~Dufresne, and C.~Jacobs-Wagner, \textit{Cell} \textbf{156}, 183 (2014).

\bibitem{Kogler}
F.~Kogler, \textit{Interactions of artificial molecular machines} (Diploma Thesis, Technical University, Berlin, 2009).


\bibitem{Mikhailov_Interface_2019} 
H.~Flechsig and A.~S.~Mikhailov, \textit{J. R. Soc. Interface} \textbf{16}, 20190244 (2019).

\bibitem{Dennison_SoftMatter_2017}
M.~Dennison, R.~Kapral, and H.~Stark, \textit{Soft Matter}, \textbf{13}, 3741 (2017).

\bibitem{Gleim_PRL_1998}
T.~Gleim, W.~Kob, and K.~Binder, \textit{Phys. Rev. Lett.} \textbf{81}, 4404 (1998).

\bibitem{Chen_JPCB_2008}
X.~Chen, I.~Weber, and R.~W.~Harrison, \textit{J. Phys. Chem. B} \textbf{112}, 12073 (2008).

\bibitem{Feig_PCCP_2019}
G.~Nawrocki, A.~Karaboga, Y.~Sugita, and M.~Feig, \textit{Phys. Chem. Chem. Phys.} \textbf{21}, 876 (2019).

\bibitem{Vendeville}
A.~Vendeville, D.~Lariviere, and E.~Fourmentin, \textit{FEMS Microbiol. Rev.} \textbf{35}, 395 (2010).

\bibitem{Heuer}
B.~Doliwa and A.~Heuer, \textit{Phys. Rev. E} \textbf{61}, 6898 (2000).

\bibitem{Berthier}
E.~Tjhung and L.~Berthier, \textit{Phys. Rev. E} \textbf{96}, 050601 (2017).


\bibitem{Ooshida}
T.~Ooshida, S.~Goto, T.~Matsumoto, and M.~Otsuki, \textit{Phys. Rev. E} \textbf{94}, 022125 (2016).

\bibitem{Guo}
M.~Guo, A.~J.~Ehrlicher, M.~H.~Jensen, M.~Renz, J.~R.~Moore, R.~D.~Goldman, J.~Lippincott-Schwartz, F.~C.~MacKintosh, and D.~A.~Weitz, \textit{Cell} \textbf{158}, 822 (2014).

\bibitem{Yasuda}
K.~Yasuda, R.~Okamoto, S.~Komura, and A.~S.~Mikhailov, \textit{EPL} \textbf{117}, 38001 (2017).


\end{thebibliography}
\end{document}